\documentclass[a4paper,10pt]{article}

\usepackage{graphicx,hyperref}
\usepackage{amsmath,amsfonts,amssymb}
\usepackage{wasysym}
\usepackage{fullpage}
\usepackage{enumerate}
\usepackage{sidecap}
\usepackage{lscape}
\usepackage{authblk}
\usepackage{appendix}
\usepackage{setspace}
\usepackage{color}

\DeclareGraphicsExtensions{.pdf}
\title{Dynamics of deceptive interactions in social networks}
\author[1,2]{Rafael A. Barrio}
\author[3]{Tzipe Govezensky}
\author[4,2]{Robin Dunbar}
\author[5,2]{Gerardo I\~{n}iguez \thanks{gerardo.iniguez@cide.edu}}
\author[2,4,6,7]{Kimmo Kaski}
\affil[1]{\small{Instituto de F{\'{\i}}sica, Universidad Nacional Aut{\'o}noma de M{\'e}xico, 01000 M{\'e}xico D.F.,
 Mexico}}
\affil[2]{Department of Computer Science, Aalto University School of Science, FI-00076 AALTO, Finland}
\affil[3]{Instituto de Investigaciones Biom{\'e}dicas, Universidad Nacional Aut{\'o}noma de M{\'e}xico, 04510 Mexico D.F., Mexico}
\affil[4]{Department of Experimental Psychology, University of Oxford, OX1 3UD, United Kingdom}
\affil[5]{Centro de Investigaci{\'o}n y Docencia Econ{\'o}micas, Consejo Nacional de Ciencia y Tecnolog{\'\i}a, 01210 M{\'e}xico D.F., Mexico}
\affil[6]{CABDyN Complexity Centre, Said Business School, University of Oxford, OX1 1HP, United Kingdom}
\affil[7]{Center for Complex Network Research (CCNR), Department of Physics, Northeastern University, Boston MA 02115, USA}

\begin{document}

\maketitle

\begin{abstract}
In this paper we examine the role of lies in human social relations by implementing some salient characteristics of deceptive interactions into an opinion formation model, so as to describe the dynamical behaviour of a social network more realistically. In this model we take into account such basic properties of social networks as the dynamics of the intensity of interactions, the influence of public opinion, and the fact that in every human interaction it might be convenient to deceive or withhold information depending on the instantaneous situation of each individual in the network. We find that lies shape the topology of social networks, especially the formation of tightly linked, small communities with loose connections between them. We also find that agents with a larger proportion of deceptive interactions are the ones that connect communities of different opinion, and in this sense they have substantial centrality in the network. We then discuss the consequences of these results for the social behaviour of humans and predict the changes that could arise due to a varying tolerance for lies in society.
\end{abstract}

\vspace{.1in}
\textbf{keywords:} social networks, anthropology, self-organised systems, mathematical modelling
\vspace{.2in}

\section{Background}
\label{sec:intro}

Deception, withholding information, making misleading statements, or blunt lies, are attitudes that most societies abhor, and parents, mentors and educators invest a great deal of effort in teaching that such behaviour is wrong and damages society~\cite{lachmann2001cost,wang2009bad,xu2010lying,popliger2011predictors}. Yet it is also true that deception and lies are present in practically all human interactions and societies~\cite{depaulo1996lying,hancock2004deception,serota2010prevalence,panasiti2011situational}. This being so, we must conclude that there is a fundamental reason that prevents the social world from being totally honest.

Broadly speaking, trust-based exchange relationships play an important role in the emergence of cooperation and complex structure in many social, economic, and biological systems~\cite{nowak2005evolution,roberts2003development,kim2009effects}. In human societies trust promotes people's willingness to engage in reciprocity~\cite{paunonen2006you}, while deception is likely to destroy the stability of such relationships by only favouring particular individuals~\cite{searcy2005evolution,sutcliffe2012computational}. However, most research has been focused on how to detect and police deception~\cite{nyberg1993varnished,craik2008reputation}, rather than on the mechanisms regulating the appearance of lies and their implications for the structure of social networks.

Previously we have studied deception and its societal consequences by means of an agent-based opinion formation model~\cite{iniguez2014effects} where the state of an agent $i$ is described with two time-dependent variables, i.e. its true (but not public) opinion $x_i$ and its public opinion $y_i$, in principle different from $x_i$. Their difference $d=|x_i-y_i|$ quantifies the lies that are told by agent $i$ to its neighbours,  which are linked by weighted links $A_{ij}$ representing social interactions. Agents and links constitute a highly structured social network where opinion formation takes place.  Both state variables evolve with a characteristic time scale $dt$, while link weights change on a different time scale $D$. In addition, the network structure co-evolves with the opinion dynamics via a network rewiring process with its own slower time scale, such that the weakest links are cut and the same number of new links are randomly created to conserve the average degree of the network. In the model, deception is defined as a partially truthful exchange of information between agents (that is, a linear combination of $x_i$ and $y_j$) with the overall proportion of honesty in the system regulated by a single parameter. Thus lies may be considered as {\it pro-} or {\it anti-social} interactions if the information passed from agent $i$ to agent $j$ is proportional to $y_j$ or $-y_j$, respectively. The selection of pro- or anti-social deception mimics the agent's intention to be as similar or different as possible from its neighbour~\cite{gino2009dishonesty}. In this context, pro-social lies are those that benefit the recipient rather than the perpetrator, for example by continuing to reinforce the dyadic relationship between them. Common examples might be `liking' something on someone's social media page even though one does not really, or asserting that something is fine when in fact it is not.

This quite simple model already gives some hints about what the fundamental utility for lying might be. We discovered that, although anti-social lies destroy the connectivity of the social network, a certain frequency of pro-social deception actually enhances the centrality of liars (who serve as links between small communities of honest people). However, in this model individuals are assumed to pursue a fixed strategy: they are always honest individuals, pro-social liars or anti-social liars. In more realistic scenarios, of course, there are enormous fluctuations away from this simple fixed strategy set and individuals vary their behaviour between the three strategies according to circumstances, even though they may naturally tend towards one strategy most of the time. An important step in exploring the dynamics of social deception, then, is to develop a model that incorporates a significant amount of strategic flexibility at the individual level. Apart from adding more realism to the model, this has the important consequence of allowing individuals and populations to evolve towards a natural equilibrium, as individuals adjust their own behaviour in accordance with the cost and benefit regimes they encounter~\cite{searcy2005evolution,szamado2000cheating,rowell2006animals}. 

The fundamental question in modelling deception is: why do people lie? In human deception, the key issue must be related to the benefits and costs when deciding what information to pass on in an interaction with another person. From this point of view, lying is a decision-making problem with an optimal solution dependent on the gains and risks of lying~\cite{gneezy2005deception,mitri2009evolution}. It is therefore important to include in the model some way of deciding the best possible answer in every instantaneous and directed dyadic interaction. In this paper we propose a more realistic model for the dynamics of deception including these features. First we describe the model in general, including the dynamics of link weights and the decision-making process. Then we discuss the results of our numerical simulations and make concluding remarks.

\section{Methods}
\label{sec:model}

Like in our earlier study~\cite{iniguez2009opinion}, the basic dynamical equation for the opinion of an agent can be written as,
\begin{equation}
\label{eq:x}
\frac{\partial x_i}{\partial t} =f_s(i)|x_i|+\alpha_if_l(i),
\end{equation}
where the state variable $x_i$ is bounded by [-1,1] and represents the instantaneous opinion of agent $i$, such that -1 corresponds to total disagreement and +1 to total agreement with a given topic of discussion. The first term on the right-hand side describes an exchange of information between a pair of agents through discussion, i.e. the interaction is short range. The second term stands for the influence of the overall opinion in the network on agent $i$, and hence the interaction is long range. Both terms evolve with a time scale $dt$ called `transaction time'. The parameter $\alpha_i$ is a random bounded variable that represents the attitude of agent $i$ to the overall opinion $f_l(i)$, being near -1 if the agent is inclined to go against the crowd and near +1 otherwise. 

In accord with our earlier model of deceptive interactions between agents~\cite{iniguez2014effects}, we define a second state variable $y_i$ corresponding to other agents' public perception of the true but private $x_i$, from which $y_i$ may differ in value if agent $i$ is not totally honest. The difference $d = |x_i - y_i|$ stands for the amount of dishonesty or the size of the lie. Hence the overall opinion $f_l(i)$ should be formed with the publicly available information (through social meetings, rumours, and news in the media) represented here by the time-dependent variable $y_i$,
\begin{equation}
\label{eq:fl}
f_l(i)=\sum_{\ell=2}^{\ell_{max}}\frac{1}{\ell}\sum_{j \in m_{\ell}(i)}y_j(t),
\end{equation}
where the second sum is over the set $m_{\ell}(i)$ of all agents $j$ separated from agent $i$ by a shortest-path length $\ell = 2,\ldots,\ell_{max}$. We assume that the influence of an agent decays with the distance $\ell$, i.e. the smallest number of links needed to reach $j$ from $i$ in the network. Without loss of generality we also consider a $1/\ell$-dependence.

In Eq.~(\ref{eq:x}) the short-range term is the direct interaction between agents  with $\ell=1$,
\begin{equation}
\label{eq:fs}
f_s(i)=\sum_{j\in m_1(i)}w_{ij}(t),
\end{equation}
where $w_{ij}(t)$ is the instantaneous information that agent $j$ passes to $i$ [see Eq.~(\ref{eq:wij})]. Observe that in general the matrix $\mathbf{w}$ is not symmetric; that is, the information that agent $i$ gives to $j$, $w_{ji}\ne w_{ij}$. Therefore, the sum of the elements of a row in $\mathbf{w}$ gives $f_s(i)$, while the sum of the elements of each column in $\mathbf{w}$ is proportional to the average apparent opinion the agent holds in the network,
\begin{equation}
\label{eq:y}
y_i(t)=\frac{1}{k_i}\sum_{j\in m_1(i)}w_{ji}(t),
\end{equation}
 where $k_i = |m_1(i)|$ is the degree of agent $i$. Explicitly, the public opinion $y_i$ is the average of the instantaneous information $w_{ji}$ received by all neighbours $j$, and is thus bounded between -1 and +1. Finally, we define the instantaneous information $w_{ij}$ as,
\begin{equation}
\label{eq:wij}
w_{ij}(t)= \phi_0(j,i,t),
\end{equation}
where the optimal opinion $\phi_0$ that agent $j$ shares with agent $i$ (i.e. between truth and pro- or anti-social lies) is the result of an individual decision-making process, as explained in section~\ref{ssec:decision}.

\begin{figure}
\centering
\includegraphics[width=0.6\columnwidth]{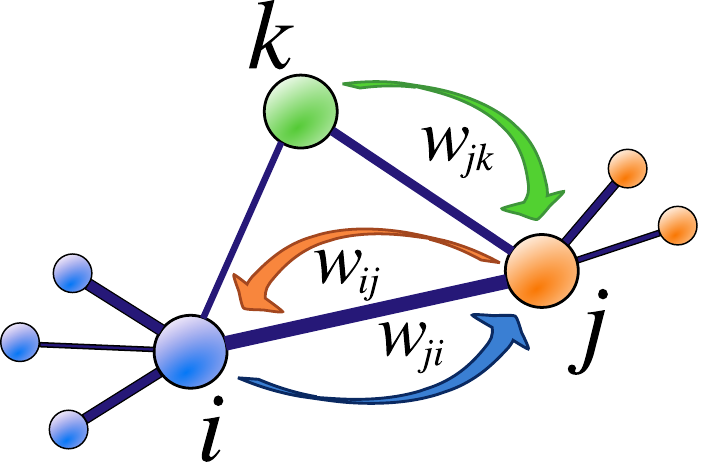}
\caption{(Online version in colour.) Diagram of the transaction dynamics of the model. Agent $i$ perceives the opinion $w_{ij}$ from neighbour $j$ and changes its own true opinion according to Eq.~(\ref{eq:x}). }
\label{fig:figure1}
\end{figure}

The nature of direct transactions is illustrated in Fig~\ref{fig:figure1}. For example, the terms $w_{ij} |x_i|$ in Eq.~(\ref{eq:x}) imply that if $w_{ij}$ has the same sign as $x_i$, agent $i$ will reinforce its position and get closer to the extreme opinion $\text{sign}(x_i)$. Next we introduce the dynamical processes involved in our deception model, as described in the sections below.

\subsection{Dynamics of link weights and rewiring scheme}
\label{ssec:links}

In social networks individuals are connected by weighted links that vary over time in the course of their dyadic interactions and decision making. We assume that `bad' decisions (not necessarily due to lies) are punished by weakening the link weight $A_{ij}$ between agents $i$ and $j$. This can be incorporated into the model by introducing a simple dynamics for link weights,
\begin{equation}
\label{eq:dA}
\frac{\partial A_{ij}}{\partial t} = DT_{ij}(t),
\end{equation}
where $D$ sets the time scale of change and $T_{ij}$ is a function of the four site variables associated with a link, namely $(x_i,y_i)$ and $(x_j,y_j)$. Since $A_{ij}$ depends on two agents, we choose the following symmetric form,
\begin{equation}
\label{eq:TP}
T_{ij}(t)= \big\{ \big| [x_i(t)+y_j(t)]+[y_i(t)+x_j(t)] \big| - 1 \big\} - P_{ij}(t),
\end{equation}
where the first square bracket represents similarity between agents according to the information agent $i$ has at its disposal, the second bracket is the corresponding term for agent $j$, and $P_{ij}(t)$ is the instant punishment for lying. Observe that the term in $\{...\}$ varies between 3 and -1, such that links with $T_{ij} < 0$ are at risk of being cut, as $A_{ij}$ approaches zero. The matrix $\mathbf{T}$ should be symmetric under exchange between $i$ and $j$, in which case the punishment the society imposes on liars reads as follows,
\begin{equation}
\label{eq:P}
P_{ij}(t)= (1-e)\left(|w_{ji}(t)-x_i(t)|+|w_{ij}(t)-x_j(t)|\right),
\end{equation}
where $e$ is a parameter that measures the tolerance of society against lies, being 0 if it is intolerant, and 1 if it does not punish liars. Thus, the punishment $P_{ij}$ is proportional to the difference between the true opinion of an agent and the instantaneous information it shares with its neighbour.

In a real social network, its topology may coevolve with the dyadic interactions between individuals~\cite{gross2009adaptive}. Thus we introduce a rewiring scheme with dynamics dependent on link weights. We proceed by cutting links with negative weight ($A_{ij}\le0\rightarrow A_{ij}\equiv0$) and immediately creating a new link (with initial weight 1) to substitute the cut link, in order to maintain the average degree of the network constant. This sets the time scale for network rewiring larger than $dt$ and variable, unlike in our former model in which rewiring is performed at fixed intervals~\cite{iniguez2014effects,iniguez2009opinion}.

The creation of new links is performed as follows. First we identify the two agents involved in the cut link and choose the one with the least number of neighbours (i.e. the most isolated agent); then we look at the second neighbours of this individual and create a link with the second neighbour (friend of a [possibly lost] friend) that has the lowest degree. This bias for favouring agents with only a few links assumes that such agents are more keen on making new friends. If there are no second neighbours, then we create a link with one of the agents with the lowest degree in the whole network. As a further remark, we note that the instantaneous information $w_{ij}$ is not necessarily the same for everyone all the time [see Eq.~(\ref{eq:wij})], the net effect of which is that the rewiring time is variable and controlled for each link by the slope $T_{ij}$.

\subsection{Decision-making process}
\label{ssec:decision}

In the \nameref{sec:intro} we state that a key issue for human deceptive interactions is related to the benefit and cost of lying, which an individual needs to evaluate in order to pass optimal information to others. This means that in each transaction, acting agent $j$ has to make a decision whether to lie or not to neighbour $i$, by finding the extremal values of a utility function $R$ that includes all gains and costs of being deceitful or truthful,
\begin{equation}
\label{eq:R}
R(\phi) = H - L(\phi) = G_H - C_H - G_L(\phi) + C_L(\phi),
\end{equation}
where $\phi$ is the opinion agent $j$ decides to share with $i$, either the truth ($\phi = x_j$) or a lie ($\phi \neq x_j$). Note that the gain $G_H$ and the cost $C_H$ of being honest do not depend on $\phi$, while the gain $G_L$ and the cost $C_L$ of being dishonest depend on the particular opinion $\phi$ that agent $j$ wishes to share. Then, the optimal opinion $\phi_0(j,i,t) = w_{ij}(t)$ is a stationary point of $R$ (either a maximum or minimum) in the permissible interval  $[-1,1]$, implicitly defined by,
\begin{equation}
\label{eq:statR}
\left. \frac{\partial R}{\partial \phi}\right |_{\phi_0}=0.
\end{equation}

Under these conditions, the decision-making process for agent $j$ is as follows. When interacting with neighbour $i$, agent $j$ finds the optimal opinion $\phi_0$ by solving Eq.~(\ref{eq:statR}). If $R(\phi_0)>0$, then agent $j$ ignores $\phi_0$ and shares its true opinion (i.e. $w_{ij} = x_j$), since being truthful is a `better' decision than not being truthful. Otherwise, agent $j$ shares the optimal opinion $\phi_0$. Note that in general $\phi_0$ stands for a lie, except for the case when $\phi_0 = x_j$. This particular case could be interpreted as a situation where an agent (that has initially decided to lie) finds that the optimal decision is to be honest.

For the decision-making process to be complete, we need to find concrete expressions for the gains and costs in Eq.~(\ref{eq:R}), based on the available sociological knowledge about interactions between individuals. The gain for being honest is considered to be `prestige' or `reputation'~\cite{panasiti2011situational}, which in our context is measured by the degree $k_j$. This is based on a previously studied sociological assumption~\cite{henrich2001evolution,david2013social}, namely that the more connected you are, the more prestige you have, which means that you are considered trustworthy. Therefore, we write the gain as,
\begin{equation}
\label{eq:gh}
G_H= \frac{k_j-\min(\{k\})}{\max(\{k\})-\min(\{k\})},
\end{equation} 
where we have normalised the degree to compare agents within and between communities.

The risk associated with being honest is proportional to the apparent disparity of opinion, as this distance increases antagonism between agents. In other words, people tend to use small, `white' lies to protect social links rather than put them at risk, since the difference in opinion corresponding to complete honesty may create tension in the relationship~\cite{dunbarmachin2014}. Then we write,
\begin{equation}
\label{eq:ch}
C_H=\frac{|x_j-y_i|}{2},
\end{equation}
which is normalised to make the gain and cost terms comparable.

If the main aim of an agent's deception is to avoid rejection by strengthening its social links, then everyday experience suggests that the gain due to lying has two components. First, the liar benefits by not `losing face', that is, by minimising the distance $\phi-y_j$ between its lie and public opinion so that the lie $\phi$ is not discovered easily. Second, the agent $j$ gains by mimicking the response $w_{ji}$ that agent $i$ is giving back, i.e. by pretending to be more similar to its peers than it is in reality. In this case we write,
\begin{equation}
\label{eq:gl}
G_L(\phi)=\frac{1}{2}\left[1-\frac{|\phi-y_j|}{2}\right]+\frac{1}{2}\left[1-\frac{|\phi-w_{ji}|}{2}\right].
\end{equation}

The risk of lying is also two-fold: agent $j$ could pass information that is similar to its true opinion ($x_j$) and risk a large mismatch of opinions. The bigger this difference, the higher the penalty (or cost) that the liar will incur from being found out~\cite{hancock2007}. Simultaneously, the agent could try to mimic an agreement with the public opinion of agent $i$, thereby risking a mismatch if agent $i$ is deceptive: the bigger the difference between the lie and public opinion, the bigger the cost the liar bears from being found out. This being so, the risk is the product of the two possibilities, 
\begin{equation}
\label{eq:cl}
C_L(\phi)=\frac{\beta}{4}|\phi-x_j| \times| \phi-y_i|.
\end{equation}
We have normalised Eqs.~(\ref{eq:gh})-(\ref{eq:gl}) such that all of them vary between zero and one. The coefficient $\beta$ in  Eq.~(\ref{eq:cl}) is a quantity that controls the relative weight of the cost of lying, which could depend on other social and cultural properties. We have examined the behaviour of the utility function $R$ and determined that $\beta=4$ balances the gains and costs between lying and being honest (see Supplementary Information).

To summarise, the dynamics of our model is highly non-linear: the elements of the adjacency matrix $\mathbf{A}$ depend on the vector $\mathbf{y}$ (the other agents' perception of an agent's true opinion), which in turn is calculated every interaction using $\mathbf{w}$. The matrix $\mathbf{w}$, in turn, is the instantaneous flow of information through each link, resulting from an agent's decision about the optimal information to pass on ($\phi_0$). Our new approach of casting transactions between agents as an optimised decision-making process constitutes a major difference from our earlier model~\cite{iniguez2014effects}. The benefit is that we now avoid predefining individuals as either pro- or anti-social liars. Nevertheless, we can still classify the lie $\phi_0$ in a binary way by comparing the distances  $|\phi_0-y_i|$ and $|\phi_0+y_i|$, being pro-social if the former is smaller and anti-social otherwise. Then, the threshold that classifies $\phi_0$ as a pro- or anti-social lie is 0, the midpoint between $\pm y_i$. We emphasise that $\phi_0=\phi_0(j,i,t)$ is a function of $j$, $i$, and $t$, obtained by finding a stationary point of $H-L$ as given in Eq.~(\ref{eq:R}). Allowing deception to vary in this way is more realistic than our previous approach of having fixed, predefined phenotypes that do not vary in their behaviour: everyday experience tells us that people do not say the same thing to everybody all the time.

\section{Results}
\label{sec:results}

Using the model described above, we perform extensive numerical simulations for networks of different sizes ($N = 100$, 200 and 500), starting with random initial conditions for the state variables $x_i$, $y_i$ and attitude parameters $\alpha_i$, and by following a simple Euler integration scheme for a sufficient number of time steps to obtain extreme opinions ($x_i = \pm 1$) for all agents. In some cases, we find a small number of agents remaining locked in an undecided state $|x_i| < 1$. This number depends on the values of the parameters $e$ (tolerance of society against lies) and $D$ (time scale for the growth of link weights), the only two parameters whose variation we consider. We can follow the time history of the process and monitor the state variables and also the amount of instantaneous lying for all agents in the system. We may also distinguish anti-social lies from pro-social ones by monitoring the optimal opinion $\phi_0(j,i,t)$. If this quantity is nearer to $y_i$, then we consider it as a pro-social lie, and if it is nearer to $-y_i$ we take it as an anti-social lie~\cite{iniguez2014effects}. As simulation results are qualitatively unaffected by network size, from now on we only consider $N=100$.

In Fig.~\ref{fig:figure2} we show typical realisations of the dynamics while keeping $D=3$ for the two extreme values of the parameter $e$. Observe that honest agents with similar opinion form small clusters (i.e. the network is assortative with respect to $x_i$~\cite{newman2003structure}), but there is also a large number of liars that serve as weak links between these tightly bonded communities and can dwell within very small communities. The effect of increasing social tolerance ($e=1$) is small, but surprisingly the relative number of liars is smaller when there is no punishment for lying. This result is in qualitative agreement with empirical observations made in schools with and without punishment~\cite{talwar2011punitive}, where the authors report that ``a punitive environment not only fosters increased dishonesty but also children's abilities to lie in order to conceal their transgressions''.

In Fig.~\ref{fig:figure3} we show the proportion of pro- and anti-social lies in the instantaneous response of each agent to all of its neighbours, for the case $e=1$ of Fig.~\ref{fig:figure2}. Observe that many agents lie pro-socially all the time and to all their neighbours. In contrast, there are very few anti-social lies and they do not persist, but instead disappear completely for some agents while for others they become intermittent. If we reduce the social tolerance for lying, anti-social behaviour disappears completely. Notice also that, despite using `ideal' conditions for the appearance of big lies ($e=1$), there are always some agents that behave in a totally honest manner all the time.

To analyse these results further, we find it convenient to quantify separately various groups of  agents. We focus our attention on those agents who are totally honest throughout the time line of the model, those who tell only pro-social lies, those who tell anti- or pro-social lies indiscriminately, and those who only lie anti-socially. Notice that for this kind of analysis to succeed, we need many realisations to obtain an average value for each case. Also, we need to look at probability distributions rather than well-defined categories, since the freedom to decide produces strategy changes in all agents. The model output suggests that agents who only lie anti-socially are very few in number, as can be seen from Fig.~\ref{fig:figure4} where we show the probability distribution of the proportion of anti-social lies for the case of zero tolerance ($e=0$) and no punishment ($e=1$). Note that social tolerance to lying has very little effect on the appearance of anti-social lies, and that most of the agents turn out to tell very few lies.
 
In Fig.~\ref{fig:figure5} we show the probability distribution of the proportion of lies per dyadic interaction, $r$, for agents who lie indiscriminately (Anti-Pro case) and for those who tell only pro-social lies (Pro case), for the two extreme values of the social parameter $e$. Explicitly, $r$ is the fraction of the total number of interactions that are lies. These results suggest that nearly 50$\%$ of the agents lie pro-socially a small amount of time ($<10\%$ of the total time). However, there are always a few agents who lie more frequently: 20$\%$ of agents lie all the time, regardless of the level of social tolerance. This result implies that it is disadvantageous to lie all the time. Fig.~\ref{fig:figure5} also suggests that the Pro- and Anti-Pro strategies are qualitatively quite similar, in the sense that many agents lie sporadically (small $r$) and a few agents ($\sim 20\%$) lie most of the time. Obviously, the relative numbers depend on the social tolerance parameter as well. An interesting observation here is that the lack of punishment for lies makes very small difference between the appearance of anti- or pro-social lies. 

Ref.~\cite{depaulo1996lying} reports a statistical study of the number and nature of lies told by a group of 144 individuals, the results of which are summarised in Table 2 therein for comparison with the results of our model.  For instance, the percentage of honest people (those who never tell lies) is 1.3\% for individuals recruited from a college population, and 8.6\% for individuals from a local community. Our results show that 2.7\% of agents are honest if there is punishment, and 3.5\% if there is not. Furthermore, the mean number of lies per day measured was roughly 2, and the mean number of social interactions 6, of which only 61\% were dyadic interactions~\cite{depaulo1996lying}. This means that  50\% of the dyadic interactions were lies. The area under the curve in Fig.~\ref{fig:figure5} (without distinguishing between pro- and anti-social lies) gives about 53\%, thus roughly agreeing with the experimental findings. In addition, we predict that the number of lies per social interaction (obtained by calculating the mean value for the amount of dishonesty or the size of the lie, $d$) is 0.38, in close agreement with the value $0.31\pm0.11$ reported in Table 2 of the experimental study~\cite{depaulo1996lying}.

We now investigate the social advantages of lying. This is done by examining network measures such as the weighted clustering coefficient  (WCC) and betweenness centrality (BC). In a weighted network or graph, WCC is the geometric average of link weights in a subgraph, where weights are normalised by the maximum weight in the network~\cite{onnela2005intensity,saramaki2007generalizations}. BT is the sum of the fraction of all-pairs shortest paths that pass through a node~\cite{brandes2008variants}. With these measures we see that liars serve as bridges between communities (Fig.~\ref{fig:figure2}); hence they sacrifice their WCC (belonging to highly connected clusters) in order to improve their BC (easy communication with all members of the network).

It is possible for a deceptive agent to increase its clustering coefficient provided it tells small lies ($d\le 0.1$), irrespective of whether these are anti- or pro-social, even in the face of social punishment. In Fig.~\ref{fig:figure6}, we show WCC averaged over 300 runs of the model for $N = 100$ agents. The conclusion is that from the perspective of clustering, there is no benefit to lying unless the lie is small. We also see that a society with total tolerance to lying does not provide liars with much advantage. However, when there is punishment, the agents who lie pro- and anti-socially have an advantage over totally honest agents, provided the number of lies is small. This can be seen in Fig.~\ref{fig:figure7}, where we show the WCC probability distribution for selected values of the proportion of lies per dyadic interaction $r$.

In conclusion, the real advantage to being a liar is that BC increases for pro-social liars, provided they tell small lies. This could be interpreted as a mechanism people use to fit into a social network better. In Fig~\ref{fig:figure8} we show the BC median taken over 300 runs of the model as a function of the size of lies, and for different groups of agents. We present the median, instead of the average, because for the form of distribution functions we have here, the median is a more robust quantity. In Fig~\ref{fig:figure8}(a) we show the case of zero tolerance ($e=0$), where only pro-social liars have an advantage over honest agents, provided their lies are very small. In Fig~\ref{fig:figure8}(b) we show the same for tolerance $e=1$.

\section{Discussion and conclusions}
\label{sec:conc}

Our model for the dynamics of lying in a social network of humans incorporates the relevant fact that the individual act of lying corresponds to a flexible, personal and instantaneous decision. Hence we have mapped this action to a decision-making problem for individuals in society so that they can adjust their behaviour to the situations they face. In contrast to our earlier model~\cite{iniguez2014effects}, where agents have fixed behavioural strategies, the present model is more realistic as the information an agent passes on (either as truth, or as a pro- or anti-social lie) is a function of the circumstances the agent encounters. In effect, we assume that agents learn and adjust their behaviour in the light of experience. In this respect, the present model lies at the opposite extreme from our previous model in that it does not assume that agents have inherited psychological predispositions to behave in a particular way. In all likelihood, of course, the real world probably lies somewhere between these two extremes. The fact that the findings from the two models are in broad agreement is therefore comforting, in that it suggests that irrespective of where reality actually lies our findings will be robust.

\begin{figure}
\begin{center}
\includegraphics[width=0.6\columnwidth]{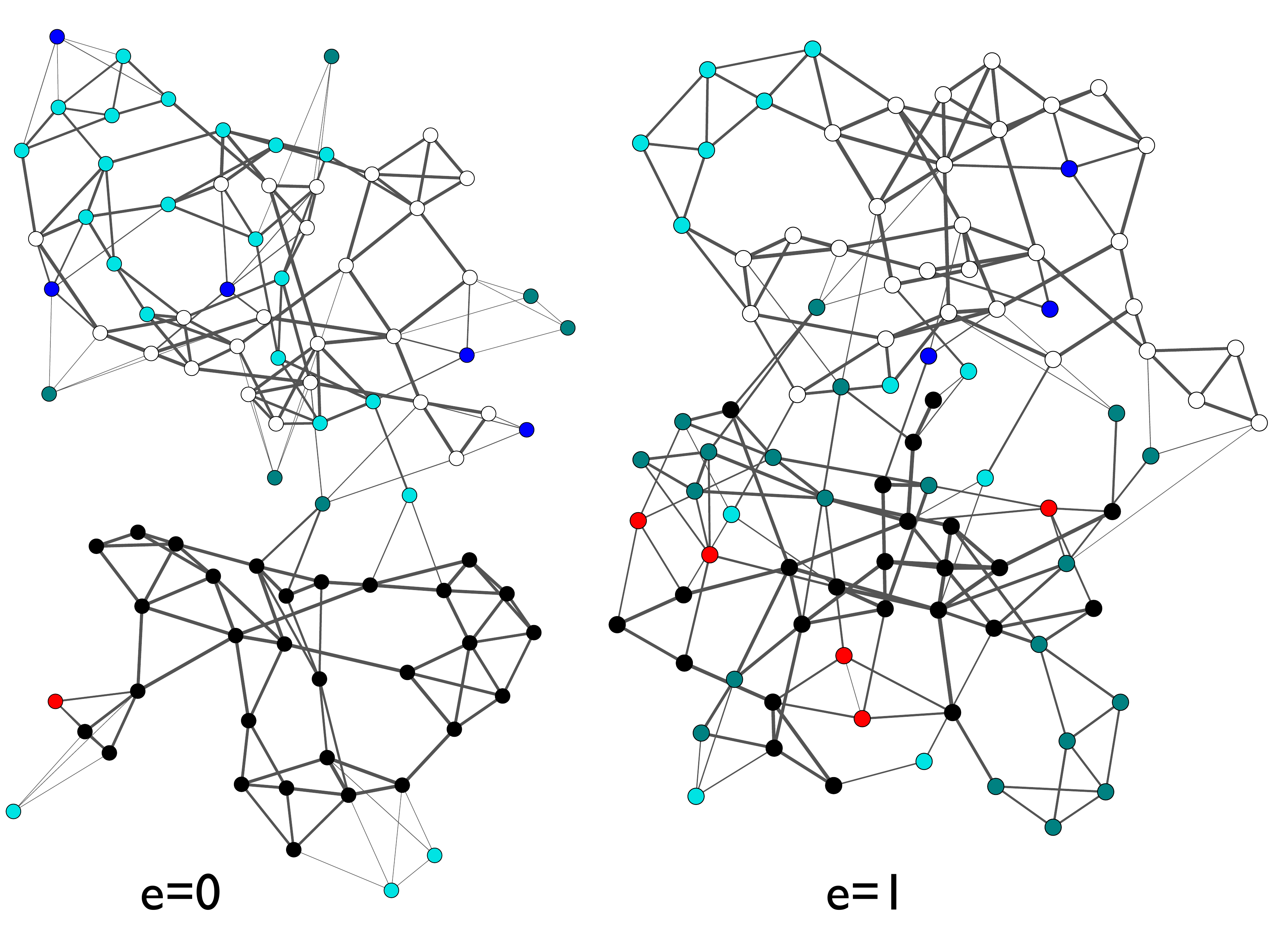}
\caption{Examples of networks with $N=100$ agents after $t_f = 600$ time iterations of the dynamics, for $e = 0,1$ and $D=3$. The color code for the agents is: honest and $x_i=1$ (white), honest and $x_i=-1$ (black), and liars (light blue or dark green) for $x_i=1$ or $x_i=-1$, respectively. The blue/red circles are undecided agents with $0 < x_i < 1$ and $-1 < x_i < 0$, respectively. Observe that the network is assortative with respect to the opinion $x_i$. The width of the links indicates their weight.}
\label{fig:figure2}
\end{center}
\end{figure}

\begin{figure}
\begin{center}
\includegraphics[width=0.6\columnwidth]{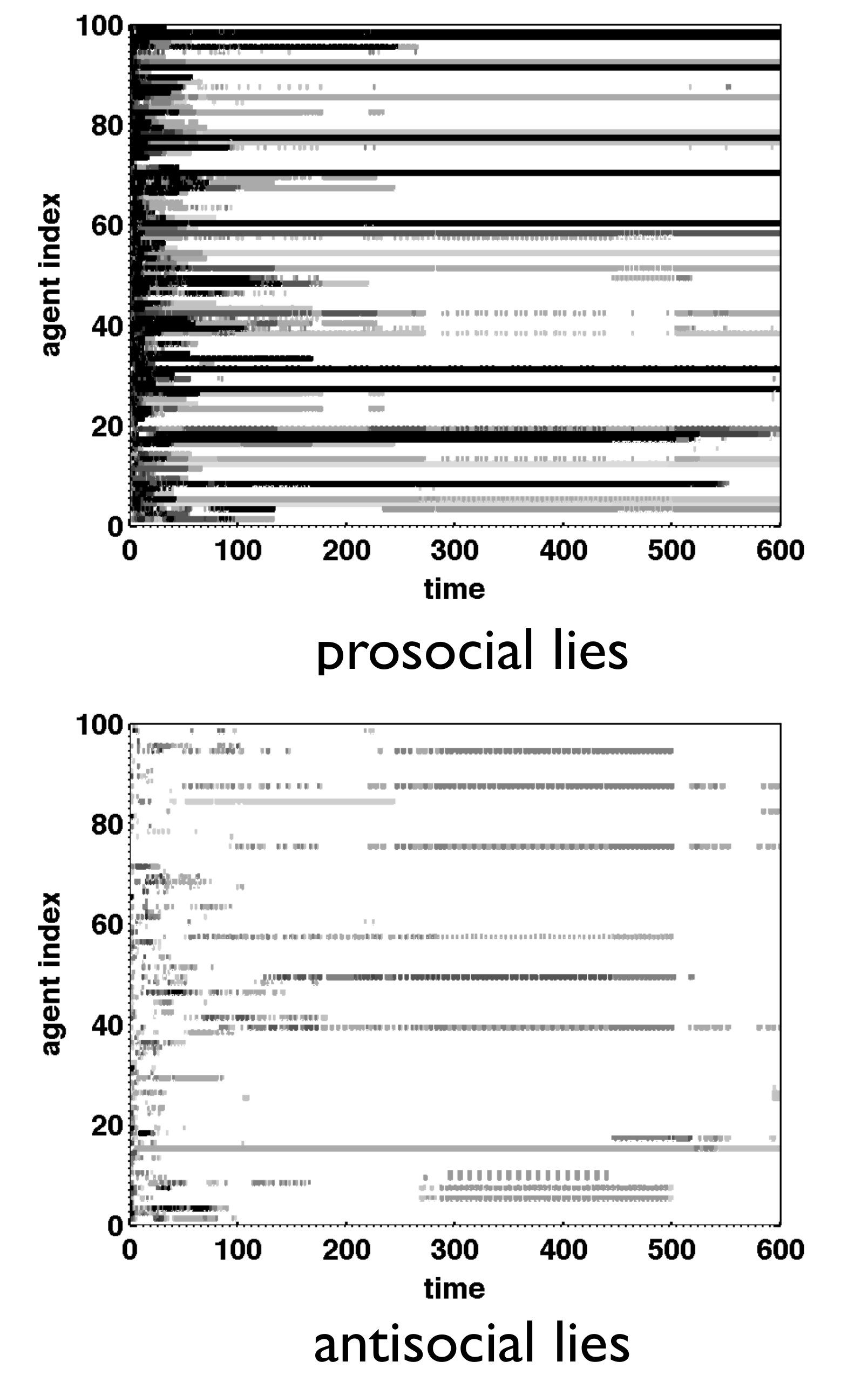}
\caption{Time history of the proportion of pro- and anti-social lies in the instantaneous response of each agent to all of its neighbours, for $e=1$. Dark/light dots correspond to high/low proportions, respectively. Observe that pro-social lies are abundant and persistent. On the contrary, anti-social lies are very few and intermittent.}
\label{fig:figure3}
\end{center}
\end{figure}

\begin{figure}
\begin{center}
\includegraphics[width=0.6\columnwidth]{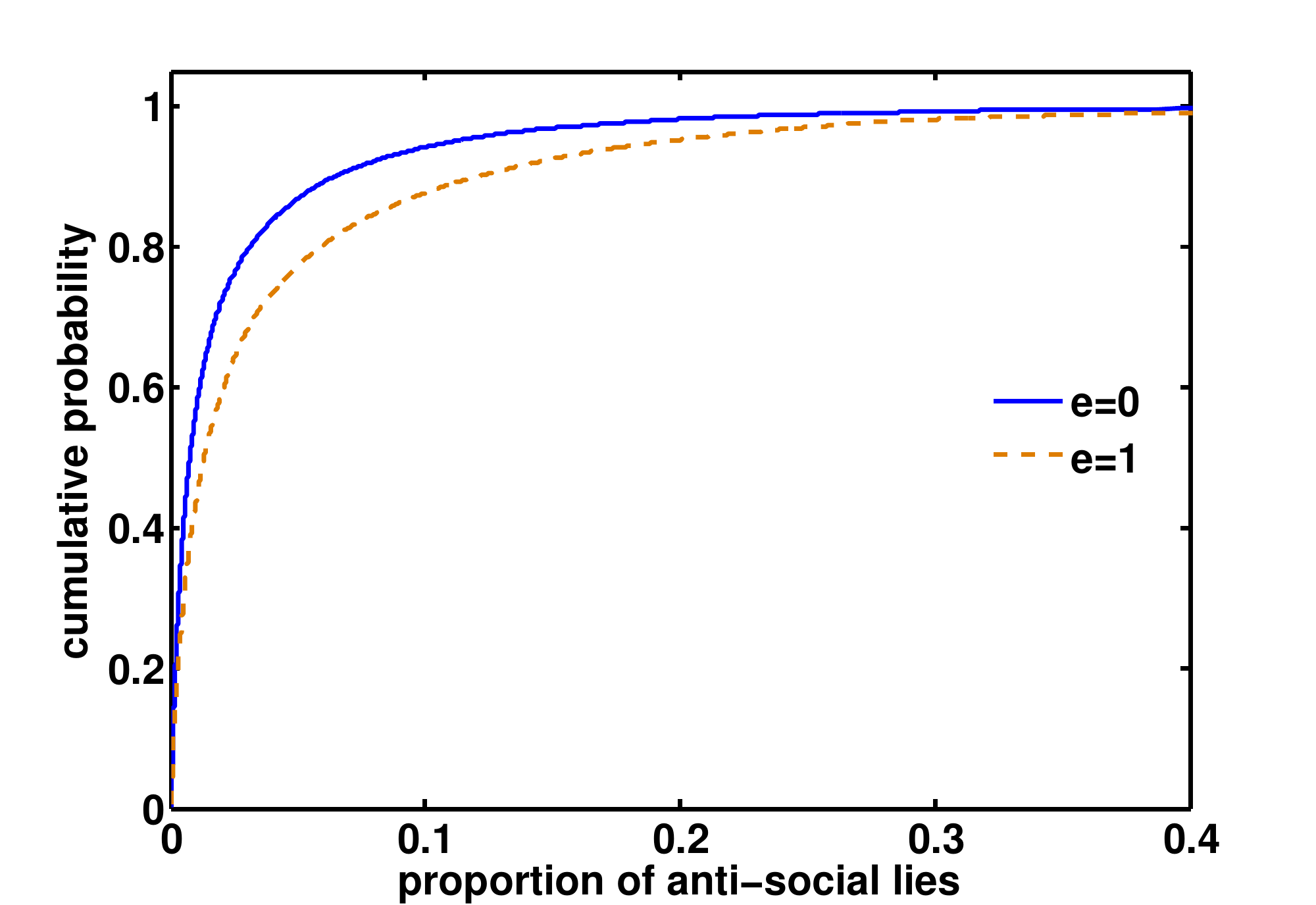}
\caption{(Online version in colour.) Probability distribution of the proportion of anti-social lies, obtained from 300 runs of networks with $N = 100$ agents. The two extreme cases of social tolerance to lies ($e = 0$, 1) are shown.}
\label{fig:figure4}
\end{center}
\end{figure}

\begin{figure}
\begin{center}
\includegraphics[width=0.6\columnwidth]{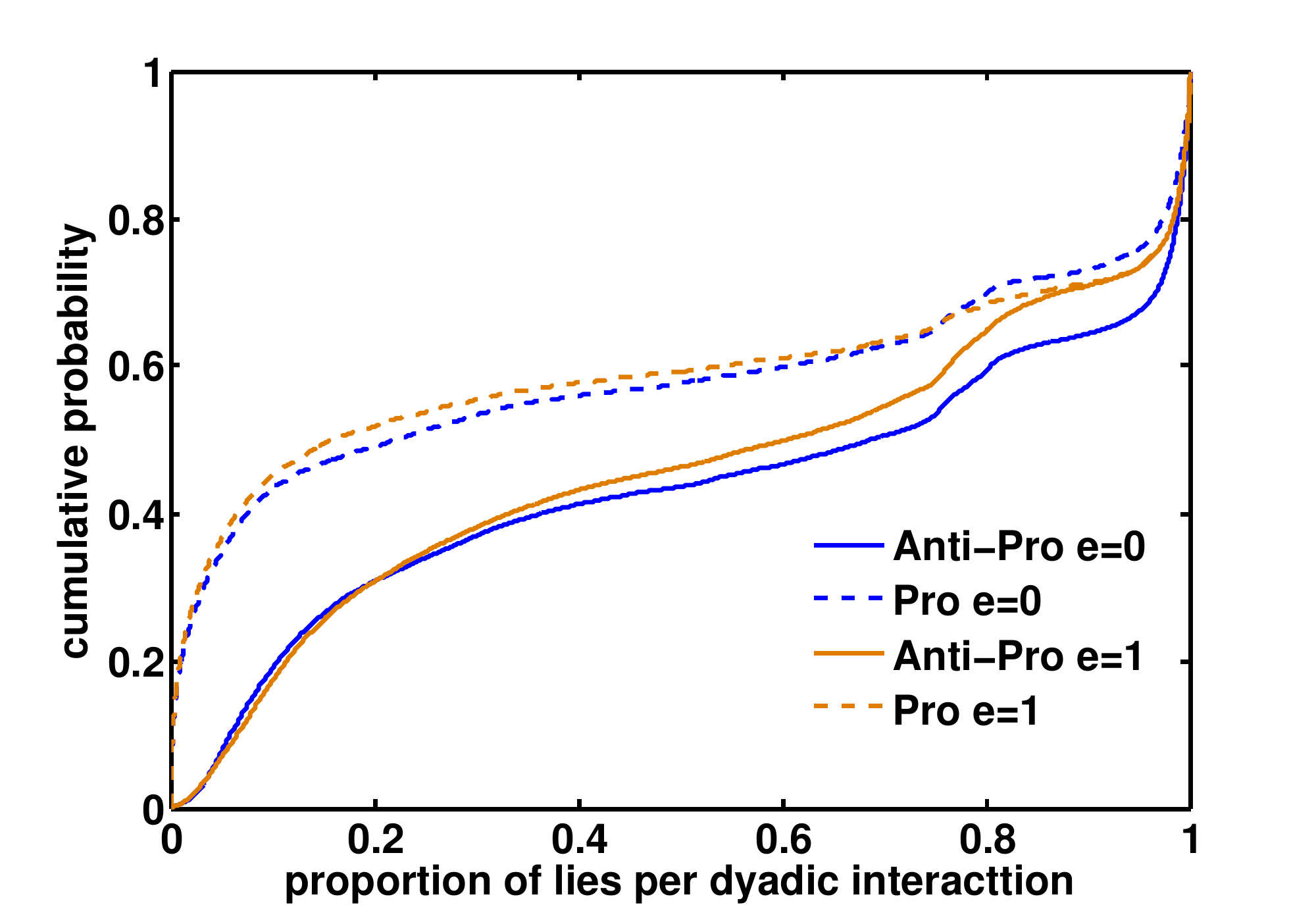}
\caption{(Online version in colour.) Probability distribution of the proportion of lies per dyadic interaction, obtained from 300 runs of networks with $N = 100$ agents. Dotted lines include pro-social lies only, while continuous lines indicate both anti- and pro-social lies. Orange/blue (light/dark gray) lines correspond to $e = 1, 0$, respectively. The normalisation factor is the total time span of each dynamics.}
\label{fig:figure5}
\end{center}
\end{figure}

\begin{figure}
\begin{center}
\includegraphics[width=0.6\columnwidth]{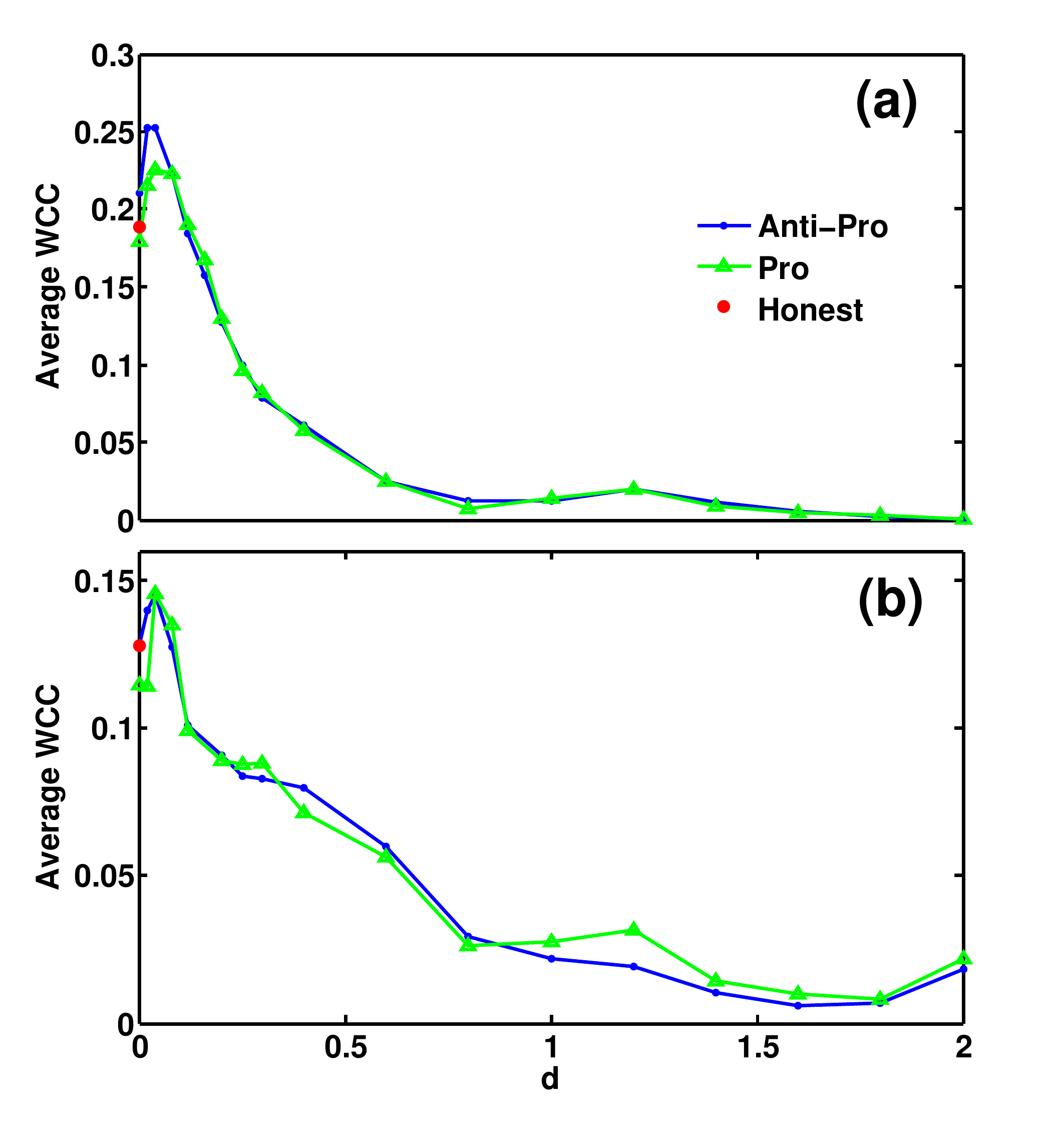}
\caption{(Online version in colour.) Average weighted clustering coefficient (WCC) as a function of the size of the lie $d$ for: (a) zero tolerance ($e = 0$), and (b) no punishment ($e = 1$). Triangles correspond to pro-social lies only, and dots to both anti- and pro-social lies. The WCC for totally honest networks is indicated by a large dot at $d = 0$.}
\label{fig:figure6}
\end{center}
\end{figure}

The model studied here does not have a network rewiring time scale that is proportional to the fundamental transaction time scale $dt$. Nevertheless, the rewiring time scale can still be tuned by using one of two parameters: $D$ (the time scale for the growth of link weights) or $e$ (the tolerance of society against lies). In addition, we see that as the tolerance parameter $e$ increases, society is more tolerant of lies and the time at which bonds are deleted increases, thus making the process slower. Furthermore, in Fig.~\ref{fig:figure2} we see that communities are much better defined when $e=0$; as a result intolerance to lies and a potential for high punishment shortens the mean life of liars, segregating the network into communities with strong internal links.

In all our simulations we find that the number of anti-social lies diminishes, while pro-social lies persist to considerable numbers throughout the dynamical progress of the system. Here we see that the social tolerance parameter $e$ has little effect on the proportion of anti-social lies, although it regulates the total number of lies. Most of the agents lie sporadically and only very few seem to lie all the time. This indicates that `true' liars are very rare in society, although they are nonetheless very special since they have large BC. We also find that liars who tell small lies ($d < 0.1$) have larger WCC. In addition, we observe that the dynamics favours the formation of cliques of purely honest agents, and that liars are usually found to be in the perimeter of cliques and connected by weak links.

\begin{figure}
\begin{center}
\includegraphics[width=0.6\columnwidth]{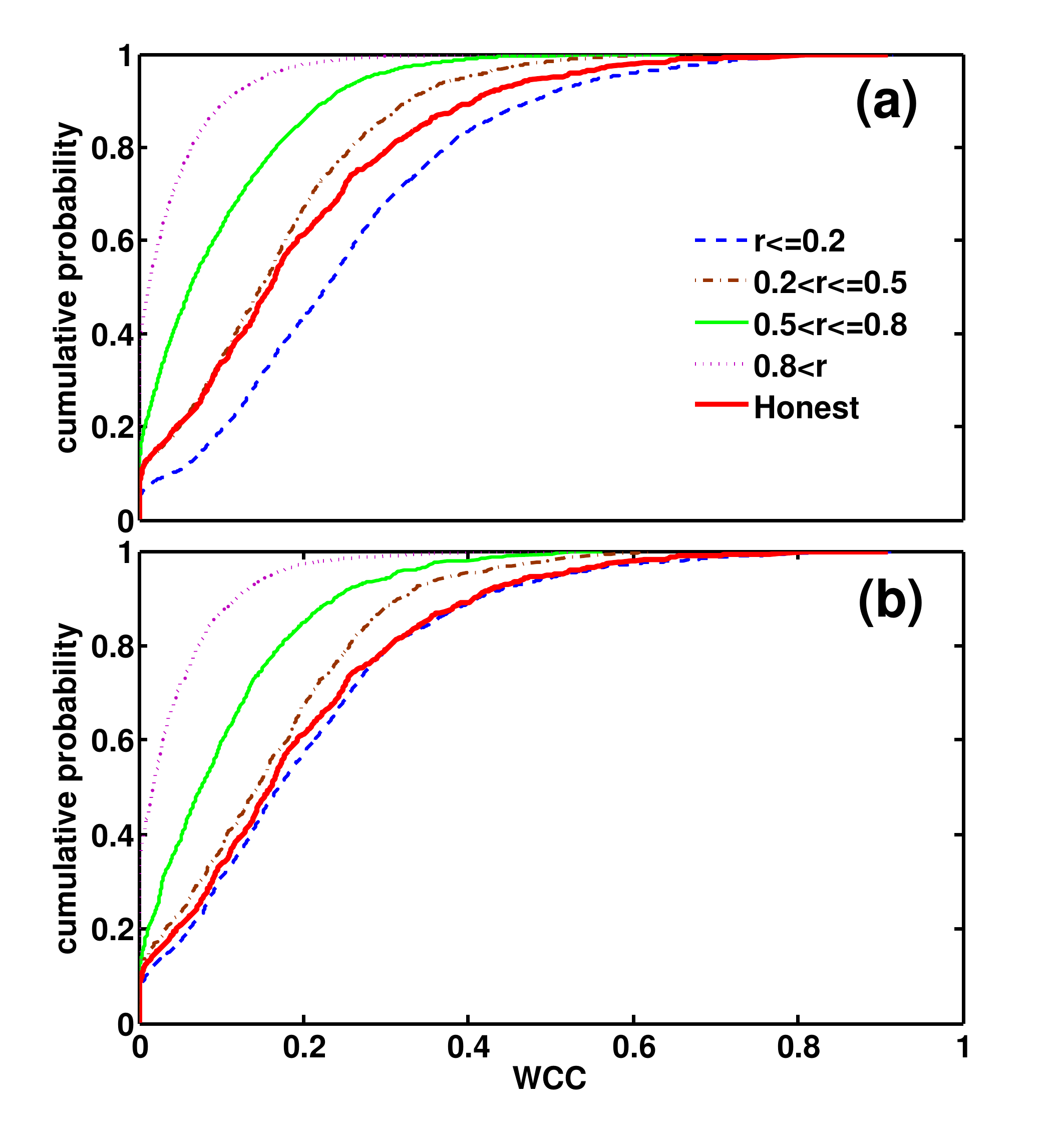}
\caption{(Online version in colour.) WCC probability distribution for several values of $r$ in a society with zero tolerance ($e = 0$), for the: (a) Anti-Pro and (b) Pro cases. Notice the advantage over honest people that agents who lie indiscriminately gain, in contrast to those who only lie pro-socially.}
\label{fig:figure7}
\end{center}
\end{figure}

\begin{figure}
\begin{center}
\includegraphics[width=0.6\columnwidth]{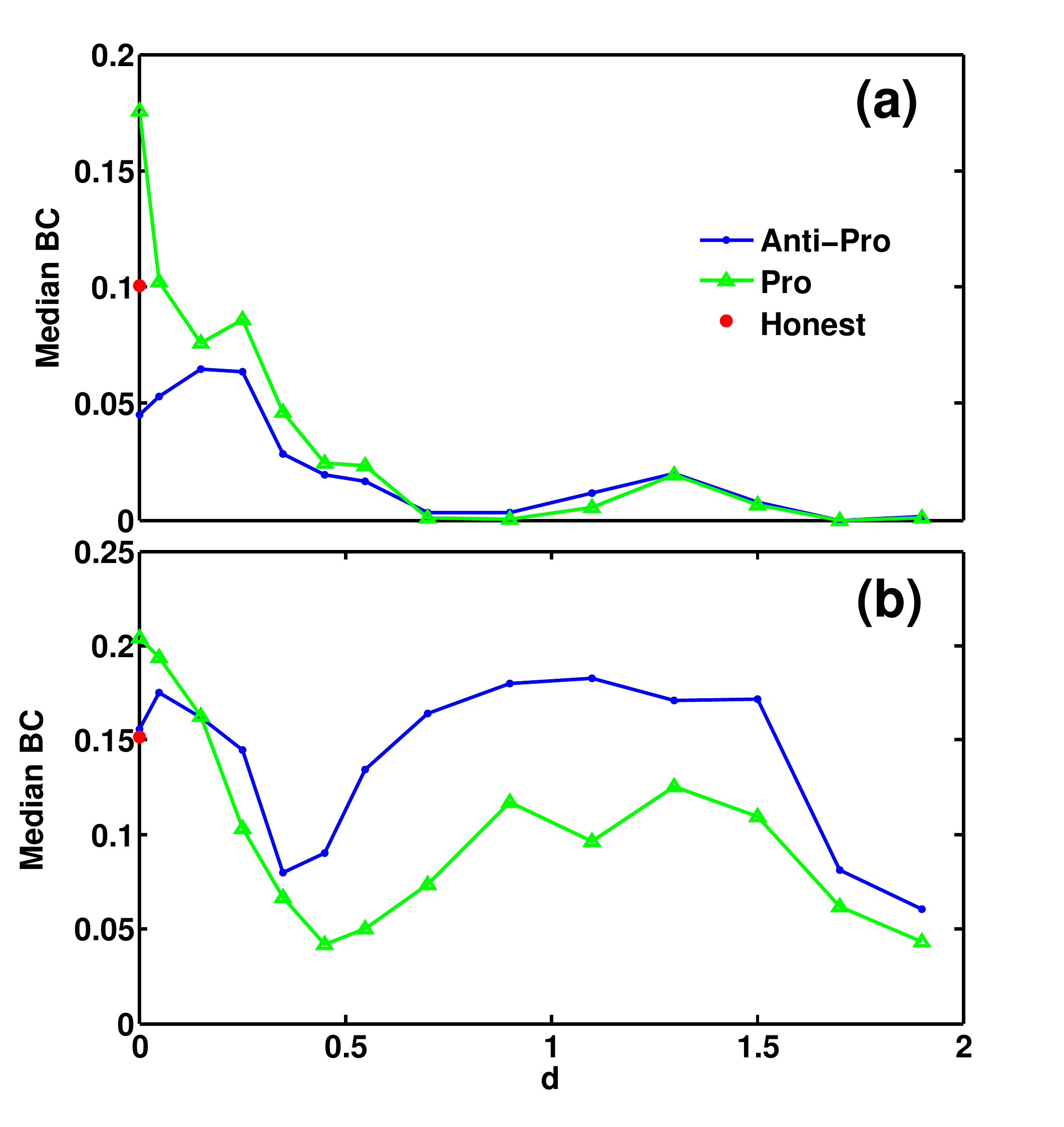}
\caption{(Online version in colour.) Median of betweenness centrality (BC), obtained from 300 runs in networks with $N = 100$ agents. Symbols are as in Fig.~\ref{fig:figure6}. Notice the advantage over honest individuals for agents who lie indiscriminately, in contrast to those who only lie pro-socially, for most lie sizes $d$.}
\label{fig:figure8}
\end{center}
\end{figure}

We also show that in general being honest pays off, but in some circumstances liars acquire an advantage over honest agents. For instance, agents who occasionally tell small lies have larger WCC and BC than honest agents (see Fig.\ref{fig:figure7} for $k< 0.2$). Moreover, an agent who tells a fair number of medium-sized lies ($d=1$) could attain a larger BC than when it chooses to be honest. 

In summary, it is interesting to note that for small lies, all liars are better off than honest agents. Even more interesting is the fact that there is a maximal advantage for people who tell sizeable anti-social lies. In short, anti-social lies yield considerable benefits for liars in appropriate circumstances. We know that anti-social lies normally destroy the social network when they are widely distributed throughout society~\cite{iniguez2014effects}. However, our findings suggest that, in certain specific circumstances, they could have the opposite effect and make the network more robust. This implies that we need to identify the conditions under which such a situation arises, by examining the local circumstances of those agents who present this peculiar property. Paradoxically, it might then be possible to increase the information flow in the network by adding appropriate motifs that allow agents to have both high BC and WCC.

\vspace{.1in}
{\bf Competing interests:} We have no competing interests.

{\bf Authors' contributions:} All authors conceived, designed and coordinated the study. RAB and GI developed and analysed the model. RAB and TG carried out the numerical and statistical analyses. All authors helped draft the manuscript and gave final approval for publication.

{\bf Funding:} RAB acknowledges support from Conacyt project No. 179616. RD's research is funded by a European Research Council Advanced grant. GI and KK acknowledge support from EU's FP7 FET Open STREP Project ICTeCollective No. 238597, and GI from the Academy of Finland.

\bibliographystyle{unsrt}
\bibliography{references}

\vspace{.2in}
\textbf{Short title for page headings:} Dynamics of deception in social networks

\newpage

\clearpage
\newpage

\begin{center}
{\LARGE Supplementary Information for}\\[0.7cm]
{\Large \textbf{ Dynamics of deceptive interactions in social networks}}\\[0.5cm]
{\large R. A. Barrio, T. Govezensky, R. Dunbar, G. I\~niguez$^*$, K. Kaski}\\[0.7cm]
{\small $^*$Corresponding author email: gerardo.iniguez@cide.edu}\\[2cm]
\end{center}

\section*{Calculation of the optimal lie}

The optimal lie $\phi_0$ is an extremal of $R$ given by Eq.~(\ref{eq:statR}), where,
\begin{align}
\label{Rapp1}
& R(\phi)=G_H-\frac{1}{2}|x_j-y_i| -\frac{1}{2}\left[\left(1-\frac{|\phi-y_j|}{2}\right)\right. \nonumber\\
&\qquad \left.+\left(1-\frac{|\phi-w_{ji}|}{2}\right)\right]+\beta\frac{|\phi-y_i||\phi-x_j|}{4},
\end{align}
according to Eqs.~(\ref{eq:R}) and~(\ref{eq:ch})-(\ref{eq:cl}). The first three terms in the right-hand side are normalised to one, while the factor $\beta$ could in principle have any value, depending on the cost we wish to apply to an act of deception. Observe that the two first terms do not depend on $\phi$. Then, taking the derivative,
\begin{align}
\label{Rapp2}
& 4\frac{\partial R}{\partial \phi}= \frac{\partial}{\partial \phi}\left[|\phi-y_j|+|\phi-w_{ji}|\right] \nonumber\\
&\, +\beta \left[|\phi-y_i|\frac{\partial (|\phi-x_j)}{\partial \phi}+|\phi-x_j|\frac{\partial (|\phi-y_i|)}{\partial \phi}\right],
\end{align}
which evaluated at $\phi_0$ [as in Eq.~(\ref{eq:statR})] gives,
\begin{align}
\label{R0}
& \mathrm{sign}(\phi_0-y_j)+ \mathrm{sign}(\phi_0-w_{ji})= \nonumber\\
&\, -\beta\left[|\phi_0-y_i| \mathrm{sign}(\phi_0-x_j) + |\phi_0-x_j| \mathrm{sign}(\phi_0-y_i)\right].
\end{align}

The solution $\phi_0$ depends on the values of the pair of signs on the left-hand side of Eq.~(\ref{R0}), denoted as $(++)$, $(+-)$, $(-+)$, or $(--)$.

\paragraph*{Case 1\\}

Let us first consider the case when the two terms on the left-hand side of Eq.~(\ref{R0}) are $(+-)$ or $(-+)$. In this situation, agent $j$ detects that there is a difference between its public opinion ($y_j$) and what agent $i$ is saying back ($w_{ji}$). This implies that,
\begin{equation}
\label{square1}
0=|\phi_0-y_i| \mathrm{sign}(\phi_0-x_j) + |\phi_0-x_j| \mathrm{sign}(\phi_0-y_i).
\end{equation}
Squaring this equation we get $(\phi_0-y_i)^2=(\phi_0-x_j)^2$, i.e.,
\begin{equation}
\label{sol1}
\phi_0=\frac{x_j+y_i}{2}.
\end{equation}
Eq.~(\ref{sol1}) is valid even for $x_j=y_i$, in which case agent $j$ should be totally honest ($\phi_0=x_j$).

\begin{figure}[t!]
\begin{center}
\includegraphics[width=0.6\columnwidth]{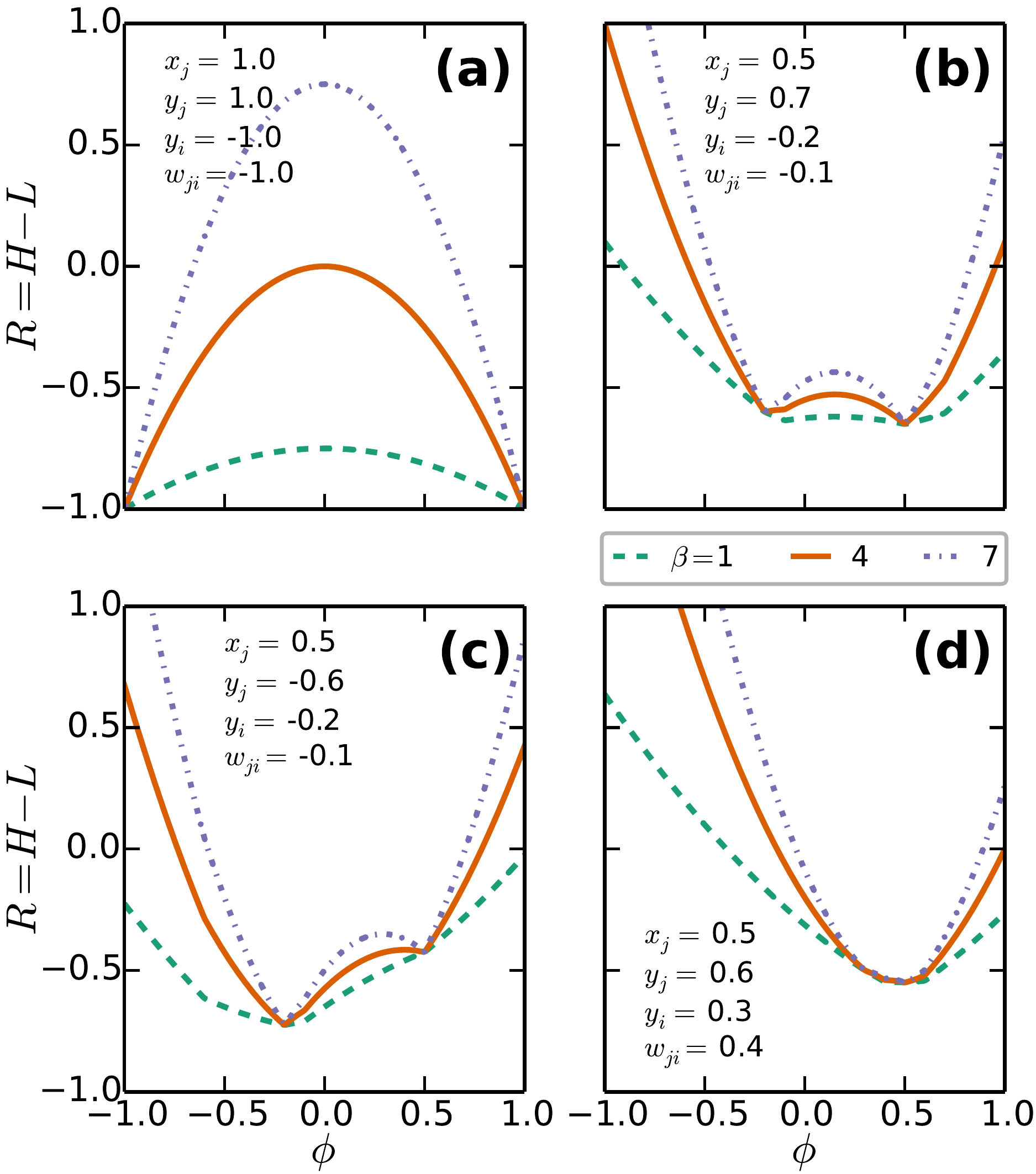}
\caption{Examples of the utility function $R(\phi) = H - L(\phi)$, for several values of the variables $x_j$, $y_j$, $y_i$, $w_{ji}$, and $\beta = 1$, 4, 7 (dashed, continuous, and dotted lines, respectively). In all cases $G_H = 0.5$.}
\label{fig:figure9}
\end{center}
\end{figure}

In other words, the lie that agent $j$ considers to be the best is exactly in the middle between its true opinion ($x_j$) and the public opinion of agent $i$ ($y_i$). This seems reasonable, since it implies an action to minimise confrontation as much as possible. Observe that in this case the solution does not depend on $\beta$.\\

\paragraph*{Case 2\\}

Now, let us consider the remaining cases $(++)$ or $(--)$ for the pair of signs in the left-hand side of Eq.~(\ref{R0}), which are only possible when agent $i$ is telling agent $j$ something very near the public opinion of $j$. From Eq.~(\ref{R0}) we have,
\begin{align}
\label{square2a}
& \pm2=-\beta\left[|\phi_0-y_i| \mathrm{sign}(\phi_0-x_j)\right. \nonumber\\
&\qquad \left.+ | \phi_0-x_j| \mathrm{sign}(\phi_0-y_i)\right].
\end{align}
Squaring this equation and considering that $\mathrm{sign}(\bullet)|\bullet|=\bullet$,
\begin{equation}
\label{square2b}
4=\beta^2\left[(\phi_0-y_i)^2+(\phi_0-x_j)^2+2(\phi_0-y_i)(\phi_0-x_j)\right],
\end{equation}
which has solution,
\begin{equation}
\label{sol2}
\phi_0=\frac{(x_j+y_i)}{2}\pm\frac{1}{\beta}.
\end{equation}

According to Eq.~(\ref{sol2}), for larger $\beta$ the lie $\phi_0$ is smaller (as expected), and for smaller $\beta$ the optimal lie is always an extreme, $\phi_0 = \pm 1$, since $\phi_0\in (-1,1)$. The sign in Eq.~(\ref{sol2}) should be chosen as to lower the value of $R(\phi_0)$.

\paragraph*{Balance in the cost of lying\\[0.1cm]}

We can use these results to estimate an appropriate value for $\beta$. The condition for this value should be that, in the case of extreme opinion values for the agents, it balances the gains and costs between lying and being honest. By extreme opinion values we mean the case when both agents have decided opposite opinions and are totally honest, i.e. $y_i=w_{ji}=\pm 1$, and $x_j=-y_i$. Since in this case the cost of lying should be greater than the cost of being honest ($C_L\ge C_H$), we have,
\[ \frac{\beta}{4} |\phi_0-x_j||\phi_0-y_i| \ge \frac{|x_j-y_i|}{2}.\]
According to Eq.~(\ref{sol1}) $\phi_0=0$ in this case, leading to,
\begin{equation}
\label{beta}
\beta \ge 4.
\end{equation}

Eq.~(\ref{beta}) sets a lower bound to the arbitrary parameter $\beta$, since for $\beta<4$ there would be situations in which telling a lie is more convenient than being honest, even against the agent's own beliefs, which we do not consider reasonable. If we interpret $\beta$ as a cultural parameter that regulates the cost of telling lies in society, then as $\beta$ increases the punishment for telling lies is large and society will tend to act more honestly as a whole. 

In Fig.~\ref{fig:figure9} we show examples of the utility function $R$ [from Eqs.~(\ref{eq:R}) and~(\ref{Rapp1})] as a function of $\phi$ for particular values of the variables $x_j$, $y_j$, $y_i$, $w_{ji}$ and $\beta$. From Fig.~\ref{fig:figure9}(a) we see that the value $\beta=4$ represents the point at which the normalized costs for lying or being honest balance, i.e. $R(\phi_0) = 0$. This means that if $\beta$ is less that 4, $R < 0 $ and agents lie all the time. Conversely, if $\beta > 4$ then agents will tend to be honest at every opportunity. In Fig.~\ref{fig:figure9}(b)-(d) we show several other possible shapes for $R$. There can be two minima with similar negative values for $R$, such that $\phi_0 = x_j$ and the optimal lie is actually the truth (b). Agent $j$ may also decide to lie if the minimum at $\phi_0 \neq x_j$ is lower (c). Finally, both agents can have similar opinions, leading to an honest interaction (d). Observe that in the last three cases, the value of $\beta$ has little effect in the stationary points of the utility function.

\end{document}